\RequirePackage{fix-cm}
\documentclass[smallextended]{svjour3}       
\smartqed  
\usepackage{graphicx,amsmath,amssymb}
\usepackage[dvipsnames]{xcolor}
\usepackage[colorlinks=true, urlcolor=blue, linkcolor=blue, citecolor=blue]{hyperref}
\begin{document}

\title{The QCD running coupling at all scales and the connection between hadron masses and $\Lambda_s$
\thanks{This material is based upon work supported by the U.S. Department of Energy, Office of Science, 
Office of Nuclear Physics under contract DE--AC05--06OR23177. }}

\titlerunning{$\alpha_s$ and the connection between hadron masses and $\Lambda_s$}        

\author{A.~Deur}

\institute{A.~Deur \at
              Thomas Jefferson National Accelerator Facility \\
              Tel.: +757-269-7526\\
              \email{deurpam@jlab.org}  
}

\date{Received: date / Accepted: date}

\maketitle

\begin{abstract}
We report on recent experimental and theoretical developments in our understanding of the QCD running 
coupling $\alpha_s$ in QCD's nonperturbative regime. They allow us to analytically compute the hadron mass 
spectrum, with $\Lambda_s$ the only input necessary to this determination. The computed spectra agrees well
with experimental data. 

\end{abstract}

\section{Introduction}
%

The strong coupling $\alpha_s$ sets the magnitude of the strong interaction and triggers the onsets of 
asymptotic freedom and confinement. As such, $\alpha_s$ is a central quantity to 
quantum chromodynamics (QCD), the gauge theory of the strong interaction.
However, while $\alpha_s$ is well understood at high energy where its smallness ($\alpha_s < 0.2$ at 
momentum transfers $Q > 10$ GeV) allows us to employ perturbative technics, it is 
much less understood at low $Q$ where $\alpha_s$ becomes large ($\alpha_s \approx 1$ at $Q \lesssim 1$ GeV). 
Thus, studying $\alpha_s$ in the strong QCD (sQCD) regime has been and remains an active field of research. Reaching 
the sQCD regime where pQCD fails is signaled by the unphysical divergence of $\alpha_s(Q)$ 
near  $Q \approx \Lambda_s$ (the Landau pole), where $\Lambda_s$ is the 
scale driving the pQCD logarithmic evolution 
of $\alpha_s(Q)$. Owing to the fact that $Q \approx \Lambda_s$ signals the breakdown of pQCD --presumably 
due to the nonperturbative confinement effects-- $\Lambda_s$ is also understood 
as the momentum scale characterizing  confinement.
We discuss here works done in the low $Q$ regime, and reported in 
Refs.~\cite{Deur:2005cf}-\cite{Deur:2017cvd}. 
Ref.~\cite{Deur:2016tte} provides a recent review of $\alpha_s$ in 
both small and large $Q$.

\section{The QCD coupling $\alpha_s$ in long distance regime \label{low Q regime}}

Different definitions of $\alpha_s$ in the sQCD domain are possible and are in fact used~\cite{Deur:2016tte}. 
In this document, the ``effective charge'' definition is employed~\cite{Grunberg:1980ja}. It defines $\alpha_s$ 
from an observable's perturbative series truncated to its first order in $\alpha_s$, and is analogous to QED's coupling 
definition (Gell-Mann Low coupling~\cite{GellMann:1954fq}).
As an example we apply below this prescription to the Bjorken sum rule~\cite{Bjorken:1966jh}, 
a fundamental relation for spin-dependent 
deep inelastic scattering. The pQCD approximant of the Bjorken sum rule is
\begin{eqnarray} 
\Gamma_1^{p-n}(Q^2)\equiv \int_{0}^{1} \big[g_1^p(x,Q^2)-g_1^n(x,Q^2)\big]dx = ~~~~~~~~~~~~~~~~ \nonumber \\ 
\frac{g_A}{6}\bigg[1-\frac{\alpha_{\overline{MS}}(Q^2)}{\pi}-
3.58\bigg(\frac{\alpha_{\overline{MS}}(Q^2)}{\pi}\bigg)^2 -\cdots \bigg] + 
\sum_{i=2}\frac{\mu_{2i}(Q^2)}{Q^{2i}},
 \label{bjorken SR}
\end{eqnarray} 
where $g_1^{p,n}(x,Q^2)$ are the longitudinal spin structure functions for the proton and 
the neutron, $x$ the Bjorken scaling variable and $g_A$ the nucleon axial charge. 
The $\mu_{2i}(Q^2)$ are nonperturbative higher-twist terms related to the confinement force~\cite{Burkardt:2008ps}. 
They become important for $Q \lesssim 1$ GeV. 
In Eq.~(\ref{bjorken SR}), the $\alpha_s$ and the series coefficients are expressed in the $\overline {MS}$ 
renormalization scheme (RS).
Using the effective charge definition, the Bjorken sum rule becomes
\begin{equation} 
\Gamma_1^{p-n}(Q^2)\equiv \int_{0}^{1}\big[ g_1^p(x,Q^2)-g_1^n(x,Q^2)\big]dx \equiv \frac{g_A}{6}\bigg[1-\frac{\alpha_{g_1}(Q^2)}{\pi}\bigg], \label{alpha_g1}
\end{equation} 
where the subscript $g_1$ for $\alpha_{g_1}(Q^2)$ indicates the observable chosen for the effective charge definition. 
This choice can be understood as equivalent to a RS choice~\cite{Deur:2016cxb}.
With this prescription, the short distance pQCD effects (the terms of second and higher orders in 
$\alpha_{\overline{MS}}$ in Eq.~(\ref{bjorken SR})) and long distance confinement 
effects ($\mu_{2i}$ terms) are folded into $\alpha_{g_1}$. This is analogous to what transmutes 
the coupling \emph{constant} in a classical lagrangian   
into a \emph{running} effective coupling, i.e. when short distance quantum effects are 
folded into the coupling definition in the renormalization process~\cite{Deur:2016tte}.
The inclusion of the long distance confinement effects removes the Landau pole of $\alpha_s$~\cite{Deur:2016tte}. 
Thus, the effective charge definition is akin to the renormalization process, with the long 
distance effects regularizing $\alpha_s$. 

There are several advantages offered by effective charges: they improve the convergence 
of the pQCD series, are extractable at any 
scale, are free of divergence and are RS-independent. This last
characteristic arises because the first order coefficient of a pQCD series is RS-independent. 
However, the price to pay for these benefits is that
an effective charge depends upon the process chosen for its definition: They are \emph{a priori} 
different effective charges for different processes.
However, QCD predictability is retained --at least in the perturbative domain-- 
since effective charges are related by commensurate scale relations~\cite{Brodsky:1994eh}. 
As already mentioned, this process-dependence is in fact equivalent to a particular choice of RS 
in the perturbative definition of $\alpha_s$.

\begin{figure}[ht!]
\centering
 \includegraphics[width=0.7\textwidth]{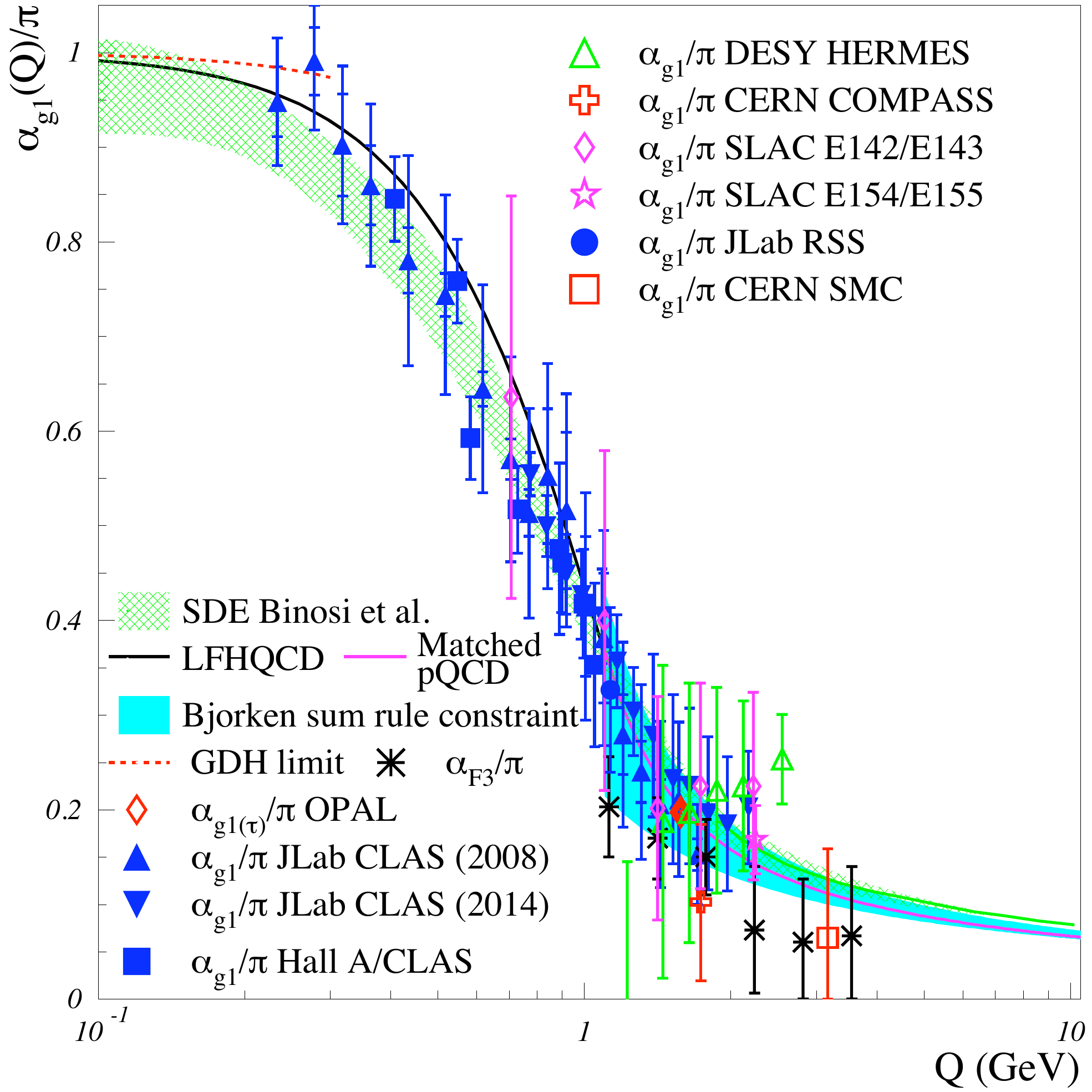}
\caption{The effective charge $\alpha_{g_1}(Q)/\pi$. The symbols indicate experimental data~\cite{Deur:2005cf}. 
The dashed red  line and solid blue band are the GDH~\cite{Gerasimov:1965et} and Bjorken~\cite{Bjorken:1966jh} 
sum rule predictions, respectively. The 
continuous black line is the LFHQCD computation~\cite{Brodsky:2010ur},  with its continuation
to the pQCD domain~\cite{Deur:2014qfa,Deur:2016opc} shown by the continuous magenta line
The green hatched band is the SDE 
calculation~\cite{Binosi:2016nme}. Comparisons with Lattice QCD calculations, older SDE results and various
models are available in Ref.~\cite{Deur:2016tte}. 
\label{fig:alpha_g1}
\protect \\
}
\end{figure}

Among the observables that can be used to define effective charges, the Bjorken sum $\Gamma_1^{p-n}$
is particularly interesting: the Bjorken sum rule at finite $Q^2$ has a relatively 
simple perturbative series, estimated up to $\alpha_{\overline{MS}}^5$~\cite{Kataev:1994gd}.
Furthermore, experimental data on $\Gamma_1^{p-n}$ exist at low, intermediate, and high $Q^2$~\cite{Adeva:1993km}. Finally, 
the rigorous Bjorken~\cite{Bjorken:1966jh} and Gerasimov-Drell-Hearn (GDH)~\cite{Gerasimov:1965et}  
sum rules dictate the behavior of $\Gamma_1^{p-n}$ in the 
unmeasured $Q^2 \to 0$ and $Q^2 \to \infty$ limits, respectively. These sum rules therefore supplement the 
data in these domains that are in practice unreachable by experiments. Consequently $\alpha_{g_1}$, the effective charge
defined using $\Gamma_1^{p-n}$,  is known at any $Q^2$. 
The experimental data supplemented by the sum rules are shown in Fig.~\ref{fig:alpha_g1}. 
The data agree well with the process-independent Schwinger-Dyson Equation (SDE) calculation 
from Ref.~\cite{Binosi:2016nme} (see also the contribution of J. Rodriguez-Quintero to these 
proceedings~\cite{Rodriguez-Quintero:2018wma}) and with the 
Light--Front Holographic QCD (LFHQCD) calculation, which will be discussed in the next Sections.
Possible reasons for why the process-depend effective charge $\alpha_{g_1}$ agrees with these process-independent 
calculations are discussed in Ref.~\cite{Deur:2016tte}. 
Finally, we note that Eq.~(\ref{alpha_g1}) and that $\Gamma_1(Q^2) \xrightarrow[Q^2 \to 0] {}  0$ impose
$\alpha_{g_1}(Q^2=0)=\pi$. $\Gamma_1$ vanishes because, as $Q^2 \to 0$, all reactions of invariant 
mass $W$ are sent to $x \to 0$. Hence the only contribution to the $\Gamma_1$ integral is at $x = 0$, 
i.e. infinite energy where cross-sections are zero.
Thus, we will refer to the $\alpha_{g_1}(0)=\pi$ relation as a kinematic constraint

\section{The Light--Front Holographic QCD approximation}
Light--Front Holographic QCD (LFHQCD)~\cite{Brodsky:2014yha}  is an approximation to QCD
based on light--front (LF) quantization~\cite{Brodsky:1997de}. This latter provides an 
exact and rigorous formulation of QCD, providing in particular a relativistic 
Schr\"{o}dinger equation that describes hadrons as quark bound-states. 
All elements of the equation --including the confining  potential term-- can be in principle obtained 
from the QCD lagrangian; 
In practice however, we know how to compute it only in (1+1) dimensions~\cite{Hornbostel:1988fb}.
The overwhelming complexity of the analytic calculations in (3+1) dimensions compels us to 
determine the potential with  methods other than first-principle computations. 
A possibility is to use the correspondence between QCD on the LF and gravity in anti-de Sitter (AdS) 
space~\cite{Brodsky:2006uqa} based on the AdS/CFT (conformal field theory) duality~\cite{Maldacena:1997re}. 
The correspondence stems from the duality between a CFT in Minkowski 
spacetime and a corresponding group of isometries of a 5-dimensional AdS spacetime. 
Thanks to this correspondence, sQCD calculations become tractable in the chiral sector
(quark masses set to zero) and if short-distance quantum fluctuations are neglected~\cite{Brodsky:2014yha}. 
The projection of the 5-dimensional AdS calculations onto the  AdS space boundary (a 4-dimensional spacetime 
identified with the physical Minkowski spacetime) provides a semiclassical approximation to sQCD that incorporates its 
fundamental aspects. In particular, it encodes by essence the  conformal
invariance of the classical QCD lagrangian, i.e. the fact that the lagrangian displays no 
explicit energy or distance scales in its expression. 

The potential of the Schr\"{o}dinger equation can be determined
using the de Alfaro, Fubini, Furlan (dAFF) procedure~\cite{deAlfaro:1976vlx}, which allows the inclusion of a scale 
in a lagrangian without explicitly breaking the conformal symmetry of the resulting action. 
The demand of explicitly preserving 
conformal symmetry in effect restricts the potential to a unique form~\cite{Brodsky:2013ar}; that of a 
harmonic oscillator on the LF,  to which is added a spin-dependent term determined by the spin representations in AdS space. Only this form yields a massless pion in the chiral sector~\cite{Dosch:2015nwa}. 
Furthermore, it explains the intriguing mass symmetry between baryons and mesons and
predicts tetraquark spectroscopy~\cite{deTeramond:2014asa,Brodsky:2016rvj}. 
Finally, a harmonic oscillator potential on the LF is equivalent in the usual instant-form 
front to the phenomenologically well-established linear potential for static quarks~\cite{Trawinski:2014msa}.
 
A single free parameter, the confinement scale $\kappa$, is used in LFHQCD. In fact, 
any theory or model describing QCD must have a least one free parameter since chiral QCD must 
be independent of conventional (human-chosen) units such as GeV. For pQCD, this single 
free parameter is $\Lambda_s$. For LFHQCD, it is $\kappa$. The relation between $\Lambda_s$ and 
$\kappa$ is analytically and numerically known~\cite{Deur:2014qfa} and its determination is a consequence 
of studying $\alpha_s$ at low $Q^2$. 
We will now show how it is determined in this 
regime using LFHQCD and how the relation between $\Lambda_s$ and $\kappa$ is obtained.

\section{LFHQCD computation of $\alpha_s(Q^2)$}
The measurement of $\alpha_{g_1}$ at low $Q^2$ and the kinematic constraint that 
$\alpha_{g_1}(Q^2=0)=\pi$ indicate that there, $\alpha_{g_1}(Q^2)$ is nearly constant, \emph{viz} 
QCD is approximately conformal at low $Q^2$. (QCD is also nearly conformal at high $Q^2$: it is 
the mildly violated Bjorken scaling).
Such behavior of $\alpha_{g_1}(Q^2)$ at low $Q^2$ could also be deduced from the GDH sum 
rule~\cite{Gerasimov:1965et} which predicts that:
\begin{equation} 
\frac{d\alpha_{g_1}(Q^2)}{dQ^2}   \xrightarrow[Q^2 \to 0] {}
 \frac{3\pi}{4g_A} \left(\frac{\kappa^2_p}{M_p^2}-\frac{\kappa^2_n}{M_n^2} \right).
\end{equation} 
where $\kappa_p=1.79$  is the proton anomalous magnetic moment, $\kappa_n=-1.91$ the neutron one, and 
$M_{p,n}$ their respective masses. That $|\kappa_n| \approx \kappa_p$ implies that 
$d\alpha_{g_1}/dQ^2  \xrightarrow[Q^2 \to 0] {} \approx 0$,
 and hence that QCD is nearly conformal at low $Q^2$.
 This fact allows us to apply LFHQCD to compute $\alpha_{g_1}(Q^2)$~\cite{Brodsky:2010ur}. 

The AdS action is similar to the Einstein-Hilbert action of General Relativity, except for being 5-dimensional: 
\begin{equation} 
S_{AdS} = -\frac{1}{4} \int \sqrt{g}\frac{1}{a^2_5}F^2~d^5x 
\end{equation} 
where $g=det(g_{\mu \nu})$; $g_{\mu \nu}$ being the AdS space metric, $a_5$ is the coupling 
in AdS space and $F$ the gauge field. 
A momentum scale (e.g. $\kappa$ or $\Lambda_s$) characterizes QCD's confining potential, which
thus breaks conformal symmetry. Hence the equivalent of the potential in Minkowski space is a distortion of its 
corresponding AdS space. As discussed in the previous section, there is a unique color confining potential, 
corresponding to a specific dilatation of AdS$_5$ space, which retains the conformal symmetry of the action, 
as in the dAFF procedure~\cite{deAlfaro:1976vlx}. This distortion takes the form of an 
exponential term $e^{\kappa^2 z^2}$ factorizing the AdS interval. Thus, the action becomes
\begin{equation} 
S_{AdS}=  -\frac{1}{4} \int \sqrt{g}\frac{1}{a^2_5}F^2 ~e^{\kappa^2 z^2}d^5x 
\end{equation} 
where $z$ is the fifth dimension of AdS space. $z^2$ gives the scale at which the hadron 
is probed, i.e. it corresponds to $1/Q^2$. The universal LFHQCD scale factor $\kappa$ is obtained  from either a hadron mass 
(e.g. $\kappa=M_\rho /2$, with $M_\rho$ the $\rho$-meson mass)~\cite{Brodsky:2014yha}, a pion or nucleon form 
factor~\cite{Brodsky:2014yha,Sufian:2016hwn} or $\Lambda_s$~\cite{Deur:2016opc}. Its value is $\kappa=0.523\pm0.024$ 
GeV~\cite{Brodsky:2016yod}.

We explained in Section~\ref{low Q regime} that, just like in pQCD in which the short distance QCD 
effects (vacuum polarization) folded into the definition
of the coupling constant $\alpha_s$ produce an effective running coupling $\alpha_s(Q^2)$, in sQCD, the 
long distance confinement forces are included into the effective charge definition. To follow this definition,
the AdS space distortion factor equivalent to the QCD confinement potential  is included in the coupling 
definition: ${a^{eff}_5}^2 \equiv a^2_5 e^{-\kappa^2 z^2}$. Transforming to the Minkowski momentum space
yields
\begin{equation} 
\alpha^{LFH}_s(Q^2) = \alpha^{LFH}_s(0) e^{-\frac{Q^2}{4\kappa^2}} \label{alpha_s from LFHQCD}
\end{equation} 
where  $\alpha^{LFH}_s(0)$ is undetermined. The kinematic constraint $\alpha_{g_1}(0)=\pi$ imposes $ \alpha^{LFH}_s(0)=\pi$ for the $g_1$ 
scheme. The factor $\alpha^{LFH}_s(0)$ thus represents the 
observable-dependence of effective charges previously discussed, which is akin to the 
RS-dependence of $\alpha_s$ in pQCD.

The LFHQCD prediction, Eq.~(\ref{alpha_s from LFHQCD}), agrees remarkably well with the 
low $Q^2$ data, see Fig.~\ref{fig:alpha_g1}, 
while having no adjustable parameters, $\kappa$ and $c$ being imposed by hadron 
masses (or form factors) and kinematic constraint, respectively. 

The expected domain of validity of the LFHQCD 
prediction is up to $Q^2 \simeq 1$ GeV$^2$. As mentioned in the previous Section, short distance 
quantum effects are not presently included in LFHQCD. Thus, LFHQCD is unsuited for high $Q^2$ phenomenology. 
Furthermore, in general AdS/CFT dualities demand the CFT coupling to be large,
which is another reason why LFHQCD is not suited for large $Q^2$ calculations.
However, $\alpha_{g_1}$ can be computed in this domain using pQCD.
In fact, the applicability domains of pQCD and LFHQCD seem to overlap around 
$Q^2 \simeq 1$ GeV$^2$~\cite{Deur:2016cxb}.
This permits us to match the $\alpha_{g_1}$ calculation from LFHQCD (denoted $\alpha^{LFH}_{g_1}$) 
to that from pQCD (denoted $\alpha^{pQCD}_{g_1}$). Specifically, 
we require  that at a scale $Q_0$,
\begin{eqnarray} 
\alpha^{LFH}_{g_1}(Q_0)=\alpha^{pQCD}_{g_1}(Q_0) \mbox{~and} \nonumber \\
\frac{d\alpha^{LFH}_{g_1}(Q)}{dQ}|_{Q=Q_0}=\frac{d\alpha^{pQCD}_{g_1}(Q)}{dQ} |_{Q=Q_0}.
 \label{eq:matching eqs.}
\end{eqnarray}  
$Q_0$ can be interpreted as the scale that sets the interface between the perturbative and nonperturbative
domains~\cite{Deur:2016cxb}. It is thus where the DGLAP~\cite{Gribov:1972ri} and ERBL~\cite{Lepage:1979zb}  
evolutions begin, see e.g. Ref.~\cite{deTeramond:2018ecg} for an application of this concept.
The existence of the domain overlap 
is validated \emph{a posteriori} by the existence of a solution to Eqs~(\ref{eq:matching eqs.}). The solution provides
both $Q_0$ and the relation between $\kappa$  and $\Lambda_s$~\cite{Deur:2014qfa}. 
Since there is a direct relation between $\kappa$ and hadron masses~\cite{Brodsky:2014yha}, this in turn provides the 
determination of the hadron mass spectrum in terms of $\Lambda_s$. The  
relation between $M_\rho$ and $\Lambda_{\overline{MS}}$ is, at leading order~\cite{Deur:2014qfa}
\begin{equation}  \label{eq: Lambda LO analytical relation}
\Lambda_{\overline{MS}}=\frac{M_\rho e^{-a}}{\sqrt{a}},
\end{equation}
where $a=4\big[\sqrt{ln^{2}(2)+\beta_{0}/4+1}-ln(2)\big]/\beta_{0}$, with $\beta_{0}=11-2n_f/3$ 
the first coefficient of the $\beta$-series of QCD. 
Would QCD be exactly conformal, the $\beta$-function (and thus $\beta_0$) would be zero and $a \to \infty$.
Eq.~(\ref{eq: Lambda LO analytical relation}) then  implies $\Lambda_s \to 0$, as expected in a conformal theory.
For actual QCD with $n_{f} = 3$ quark flavors, $a\simeq 0.55$ at LO. 
At N$^3$LO the relation must be solved numerically and is $\Lambda_{\overline{MS}}=0.440 M_\rho$. 

\begin{figure*}[ht!]
\centering
\centerline{\includegraphics[width=.49\textwidth]{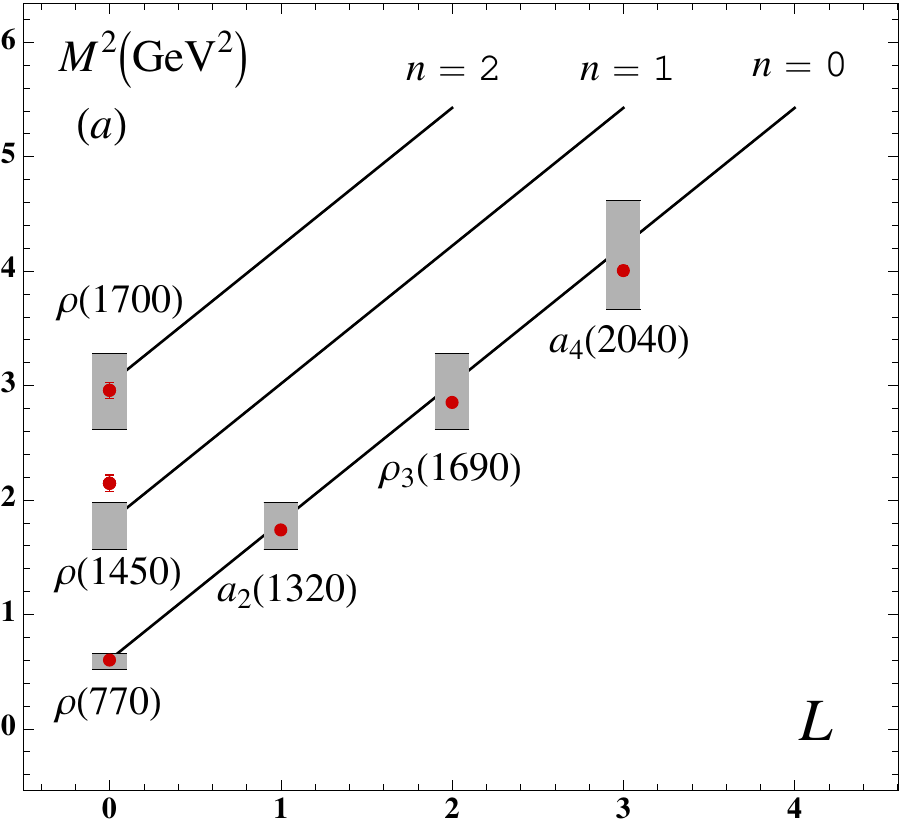} 
\includegraphics[width=.49\textwidth]{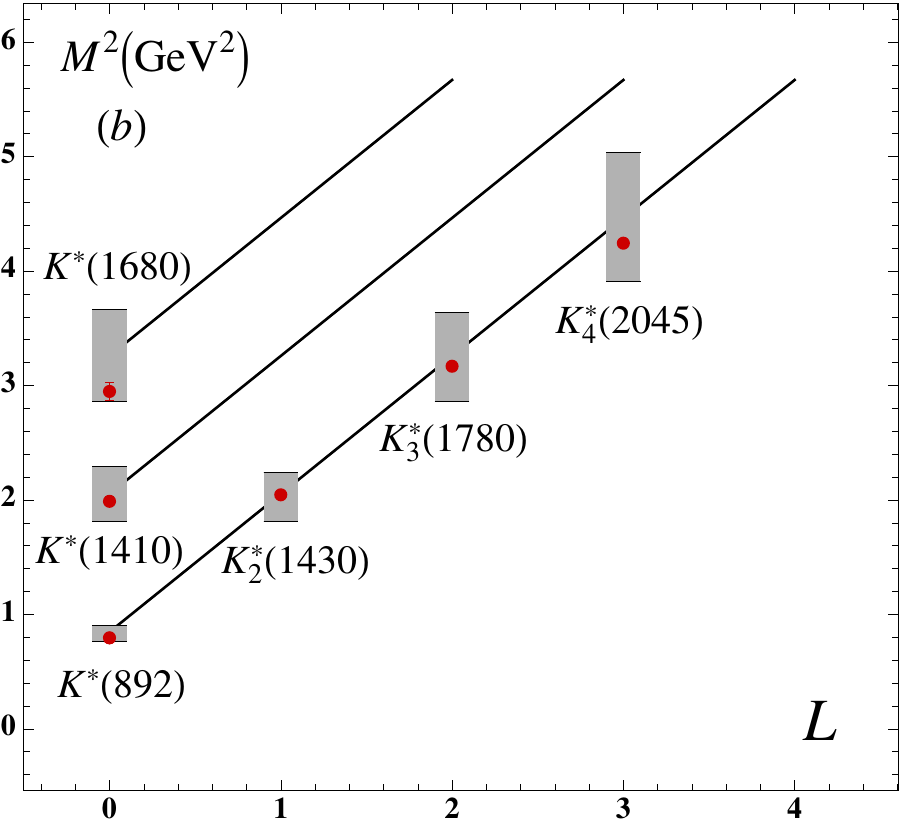}}
\caption{\label{Fig:masses} The LFHQCD prediction of the mass spectrum for unflavored (left) and 
strange (right) light vector mesons. 
The gray rectangles indicate the uncertainty on the LFHQCD calculation. The red  points are the experimental data.}
\end{figure*}

The $\rho$ meson state is the solution of the LF Schr\"{o}dinger equation with internal 
orbital angular momentum $L=0$ and radial excitation $n=0$. Higher mass mesons of the same family are 
solutions with $L> 0$ or/and $n > 0$. They can thus be obtained following the same method. The result at N$^3$LO is 
shown in Fig.~\ref{Fig:masses} together with the prediction for strange mesons obtained similarly.
Baryonic masses can be obtained likewise, or using the mass symmetry between  $L_B$ baryons and 
mesons with $L_M=L_B+1$ predicted by the superconformal algebraic structure in LFHQCD~\cite{deTeramond:2014asa}.
Hence, the method presented here provides an analytic determination of hadron spectrum 
with $\Lambda_s$ as the single input.
The method can be reversed, using the known value of $\kappa$ to predict $\Lambda_s$. This yields 
$\Lambda_{\overline{MS}}^{n_f=3}=0.339(19)$ GeV, in excellent agreement with the Particle Data 
Group average of 0.332(17) GeV~\cite{Olive:2016xmw}.

\section{Summary and conclusion}

The coupling $\alpha_s(Q^2)$ is a central element of QCD. Vigorous research efforts 
aim to understand it in the strong QCD regime~\cite{Deur:2016tte}.
One such endeavor uses the Bjorken sum rule~\cite{Bjorken:1966jh} to define an effective 
charge $\alpha_{g_1}(Q^2)$~\cite{Grunberg:1980ja}. 
Such choice is advantageous because data~\cite{Adeva:1993km} 
and sum rules~\cite{Bjorken:1966jh,Gerasimov:1965et} allow us to obtain $\alpha_{g_1}(Q^2)$ over all 
$Q^2$~\cite{Deur:2005cf}. Both the experimental data and the sum rules independently indicate that QCD is nearly conformal 
at low $Q^2$. This makes possible to compute strong QCD using the AdS/CFT correspondence~\cite{Maldacena:1997re}, in 
particular in its LFHQCD incarnation~\cite{Brodsky:2006uqa}. 
This  provides a semiclassical analytic approach to strong QCD that is fully determined and can 
be used to solve QCD on the light-front~\cite{Brodsky:2014yha}. The potential arising in the (rigorous) light-front bound-state 
equation is imposed by explicitly respecting QCD's conformal symmetry~\cite{deAlfaro:1976vlx}. Only a single free parameter, 
$\kappa$, is needed, which is in fact the minimal amount of parameters necessary for any theory or 
model aiming to describe the strong interaction. 
The strong coupling $\alpha_{g_1}(Q^2)$ obtained with LFHQCD~\cite{Brodsky:2010ur} has no adjustable parameter 
and is in remarkable agreement with the experimental data~\cite{Deur:2005cf} and 
with calculations form different approaches to strong QCD~\cite{Deur:2016tte}, including 
the recent Schwinger-Dyson Equation calculation of a process independent 
coupling~\cite{Binosi:2016nme,Rodriguez-Quintero:2018wma}. 
The fact that the validity domains of LFHQCD and pQCD 
overlap around $Q^2=1$ GeV$^2$ enables us to obtain  both the scale $Q_0$ that 
sets the interface between the nonperturbative and perturbative
domains~\cite{Deur:2016cxb,deTeramond:2018ecg} and an analytic determination of the hadron mass spectrum from 
the fundamental QCD parameter $\Lambda_s$~\cite{Deur:2014qfa}. Conversely, it permits a high-accuracy determination of 
$\Lambda_s$~\cite{Deur:2016cxb,Deur:2016opc} that agrees well with the world average~\cite{Olive:2016xmw}.

Being able to analytically compute the hadron mass spectrum has been a long thought goal 
and would signal that we finally managed to analytically solved QCD. While LFHQCD is not QCD but a 
semi-classical approximation of it, its premises are based on the rigorous LF approach to quantum field theory and the 
formal relation between the isometry group of a 5-dimensional AdS space and the conformal invariance 
of the dual theory in the 4-dimensional physical space. The foundations of LFHQCD are thus solid and 
the analytic computation of the hadron spectrum is an encouraging progress toward the ultimate goal of
analytically solving QCD.

\begin{acknowledgements}
This work was done in collaboration with S. J. Brodsky, V. Burkert, J-P Chen, G. de T\'eramond and W. Korsch.
We thank S. J. Brodsky and G. de T\'eramond 
for reading the manuscript and providing useful comments.
\end{acknowledgements}

\end{document}